\documentclass[pra,onecolumn,floatfix,a4paper,superscriptaddress]{revtex4}
\usepackage{bm,color,graphicx,amsmath,txfonts}

\begin{document}

\title{Macroscopic entanglement of two magnon modes via quantum correlated microwave fields}

\author{Mei Yu}
\affiliation{Zhejiang Province Key Laboratory of Quantum Technology and Device, 
Department of Physics and State Key Laboratory of Modern Optical Instrumentation, Zhejiang University, Hangzhou, Zhejiang, China}
\author{Shi-Yao Zhu}
\affiliation{Zhejiang Province Key Laboratory of Quantum Technology and Device, 
Department of Physics and State Key Laboratory of Modern Optical Instrumentation, Zhejiang University, Hangzhou, Zhejiang, China}
\author{Jie Li}
\affiliation{Zhejiang Province Key Laboratory of Quantum Technology and Device, 
Department of Physics and State Key Laboratory of Modern Optical Instrumentation, Zhejiang University, Hangzhou, Zhejiang, China}

\begin{abstract}
We present a scheme to entangle two magnon modes in two macroscopic yttrium-iron-garnet spheres. The two spheres are placed inside two microwave cavities, which are driven by a two-mode squeezed microwave field. By using the linear state-swap interaction between the cavity and the magnon mode in each cavity, the quantum correlation of the two driving fields is with high efficiency transferred to the two magnon modes. Considerable entanglement could be achieved under experimentally achievable conditions $g \gg \kappa_a \gg \kappa_m$, where $g$ is the cavity-magnon coupling rate and $\kappa_a$, $\kappa_m$ are the decay rates of the cavity and magnon modes, respectively. The entanglement is in the steady state and robust against temperature, surviving up to hundreds of milliKelvin with experimentally accessible two-mode squeezed source.

\end{abstract}

\date{\today}

\maketitle

\section{Introduction}

In recent years, ferrimagnetic systems, like yttrium iron garnet (YIG), become an active and important platform for the study of strong interaction between light and matter, owing to their high spin density and low damping rate. Magnons, as collective excitations of a large number of spins, can strongly couple to cavity microwave photons~\cite{Strong1,Strong2,Strong3,Strong4,Strong5,Strong6,Tobar2,Tobar3} leading to cavity-magnon polaritons. The strong coherent interaction allows one to observe many interesting phenomena in cavity-magnon systems, such as the exceptional point~\cite{You17NC}, remote manipulation of spin currents~\cite{Bai}, bistability~\cite{You18PRL}, etc.

An intriguing direction would be hybrid systems based on magnonics~\cite{NakaRev}. The coupling of magnons with a variety of different systems, either continuous variable or discrete variable, provides great opportunities for the study of many interesting topics: e.g., coupling magnons to a superconducting qubit~\cite{NakaSci15} allows one to resolve magnon number states~\cite{Naka17SA}, coupling magnons to both optical and microwave photons~\cite{Tang16PRL} allows one to coherently convert optical and microwave photons~\cite{Naka16PRB}, and coupling magnons to the vibrational mode of a YIG sphere~\cite{Tang16SA} allows one to prepare magnon-photon-phonon entangled states~\cite{Jie18PRL} and magnon/phonon squeezed states~\cite{Jie19PRA}. In the last case, magnons and phonons are coupled via nonlinear magnetostrictive interaction, which is of radiation pressure type, and therefore many known results in the well-developed field of cavity optomechanics~\cite{omRMP} are expected to occur in the new field of cavity magnomechanics, such as magnomechanically induced transparency~\cite{Tang16SA}, and magnomechanical cooling and entanglement~\cite{Jie18PRL}. Many other interesting phenomena have been explored in cavity magnomechanics, such as slow light effect~\cite{Wu19}, phonon lasing~\cite{Li19}, and parity-time-related phenomena~\cite{Liu19,Sun19}.

In this paper, we focus on quantum effect in cavity-magnon systems and present a scheme to entangle two magnon modes in two macroscopic YIG spheres. Recently, several proposals have been put forward, using different mechanisms, for preparing entangled states of two magnon modes, either in ferrimagnetic YIG spheres~\cite{Jie19,Zhedong,Jaya} or in an antiferromagnetic system~\cite{Yung}. The two spheres are placed inside two microwave cavities, and in each cavity a magnon mode couples to a cavity mode via linear beamsplitter interaction. It is well known that such interaction will not generate any entanglement. One approach is to introduce necessary nonlinearity into the system, either from the magnetostrictive interaction~\cite{Jie19} or from the Kerr effect~\cite{Zhedong}. Another approach is to inject external quantum resource, e.g., squeezed vacuum field~\cite{Jaya}, into the system. The present scheme follows the latter approach: we drive the two cavities with a two-mode squeezed vacuum microwave field. The idea is to transfer the quantum correlation shared by the two driving fields to the two magnon modes via the linear cavity-magnon coupling. We show that this quantum state transfer occurs with high efficiency provided that the cavity and magnon modes are resonant with the driving fields, and the coupling rate $g$ and the cavity (magnon) decay rate $\kappa_a$ ($\kappa_m$) satisfy $g\gg \kappa_a \gg \kappa_m$, which has been realized, e.g., in the experiments~\cite{Strong2,Strong3,Strong6}. The two magnon modes are entangled in the steady state, and the entanglement increases with the squeezing of the input two-mode squeezed field and survives up to hundreds of milliKelvin with experimentally accessible two-mode squeezed source. 

The remainder of the paper is constructed as follows. In Sec.~\ref{Model}, we introduce the system, provide its Hamiltonian and the corresponding quantum Langevin equations (QLEs), and show in detail how to solve the QLEs and calculate the entanglement. In Sec.~\ref{Numeri}, we provide numerical results and show optimal parameter regimes where large magnon entanglement can be obtained, and in Sec.~\ref{Analy}, we provide analytical solutions at the optimal conditions and analyse in more depth the topic. Finally, we draw our conclusions in Sec.~\ref{Conc}.

\begin{figure}[t]\label{fig1}
\hskip-2.0cm\includegraphics[width=0.4\linewidth]{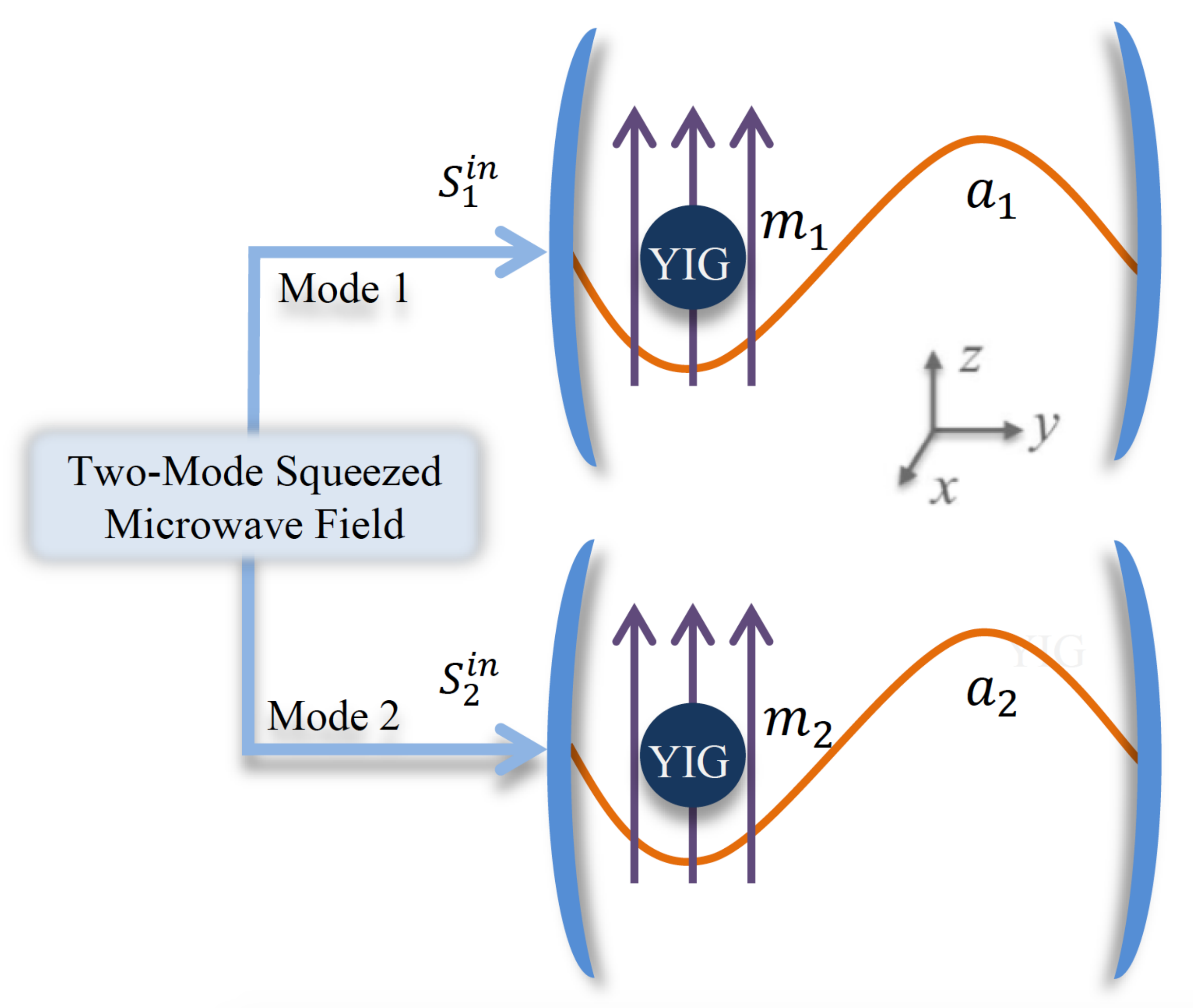} 
\caption{Sketch of the system. A double cavity-magnon system where the two cavities are driven by a two-mode squeezed vacuum microwave field. Two YIG spheres are respectively placed inside the cavities near the maximum magnetic fields of the cavity modes and simultaneously in uniform bias magnetic fields. The bias magnetic fields and the magnetic fields of the cavity modes are along the $z$ and $x$ directions, respectively.}
\end{figure}

\section{The Model}\label{Model}

The system under study, sketched in Fig. 1, consists of two microwave cavities and two magnon modes in two YIG spheres, which are respectively placed inside the cavities near the maximum magnetic fields of the cavity modes and simultaneously in uniform bias magnetic fields, which excite the magnon modes in the spheres and couple them to the cavity modes. The magnons are quasiparticles and are embodied by a collective excitation of a large number of spins in YIG spheres. The system of two YIG spheres has been used to study magnon dark modes~\cite{Tang15NC}, high-order exceptional points~\cite{You19PRB}, and entanglement properties between two magnon modes~\cite{Jie19,Zhedong,Jaya}. In each cavity, the magnon mode couples to the cavity mode via the magnetic dipole interaction, and this coupling can be very strong~\cite{Strong1,Strong2,Strong3,Strong4,Strong5,Strong6,Tobar2,Tobar3} thanks to the high spin density of YIG. The two cavities are driven by a two-mode squeezed vacuum microwave field. We consider the size of the YIG spheres is much smaller than the microwave wavelengths, and hence the effect of radiation pressure can be safely neglected. The Hamiltonian of the system reads 
\begin{equation}
{\cal H}/\hbar =\underset{j=1,2}{\sum }\Big\{ \omega _{a_{j}}a_{j}^{\dag }a_{j}+\omega_{m_{j}}m_{j}^{\dag }m_{j}+g_{j} \big( a_{j}m_{j}^{\dag }+a_{j}^{\dag }m_{j} \big) \Big\},
\label{Hamilt}
\end{equation}
where $a_{j}$ ($a_{j}^{\dag }$) and $m_{j}$ ($m_{j}^{\dag }$) are annihilation (creation) operators of the $j$th cavity and magnon modes, respectively, and we have $\big[ O,O^{\dag }\big] =1$ ($O\,{=}\,a_j, m_j$).  $\omega _{a_{j}}$ ($\omega _{m_{j}}$) is the resonance frequency of the $j$th cavity mode (magnon mode). The frequency of the magnon mode $\omega _{m_{j}}$ is determined by the external bias magnetic field $H_j$ and the gyromagnetic ratio $\gamma$ via $\omega _{m_{j}}=\gamma H_j$, and thus can be flexibly adjusted, and  $g_{j}$ is the coupling rate between the $j$th\ cavity and magnon modes.

In the frame rotating at the frequency $\omega _{_{j}}$, i.e., the frequency of the $j$th mode of the input two-mode squeezed field, the QLEs of the system are given by
\begin{eqnarray} \label{QLEs}
\overset{\cdot }{a}_{j} &=&-(\kappa _{a_{j}}+i\Delta_{a_{j}})a_{j}-ig_{j}m_{j}+\sqrt{2\kappa _{a_{j}}}S_{j}^{in},  \\
\overset{\cdot }{m}_{j} &=&-(\kappa _{m_{j}}+i\Delta_{m_{j}})m_{j}-ig_{j}a_{j}+\sqrt{2\kappa _{m_{j}}}m_{j}^{in},  \notag
\end{eqnarray}
where $\kappa _{a_{j}}$ ($\kappa _{m_{j}}$) is the decay rate of the $j$th cavity mode (magnon mode), $\Delta _{a_{j}}=\omega_{a_{j}}-\omega _{_{j}},\Delta _{m_{j}}=\omega _{m_{j}}-\omega _{_{j}}$, and $S_{j}^{in}$ ($m_{j}^{in}$) is the input noise operator for the $j$th cavity mode (magnon mode). The two cavity input noise operators $S_j^{in}$ are quantum correlated due to the injection of the two-mode squeezed field, and have the following correlations in time domain 
\begin{eqnarray}
\left\langle   S_{j}^{in\dag}(t)S_{j}^{in}(t^{\prime }) ; S_{j}^{in}(t)S_{j}^{in\dag }(t^{\prime }) \right\rangle &=&( N ; N+1) \delta(t-t^{\prime}),\,\,\,\, (j=1,2)  \label{NoiseCF1} \\
\left\langle S_{j}^{in}(t)S_{k}^{in}(t^{\prime }) ;  S_{j}^{in\dag}(t)S_{k}^{in\dag }(t^{\prime }) \right\rangle &=&( M;M^{\ast })
\delta (t-t^{\prime }),\,\,\,\,  (j\neq k=1,2)  \notag
\end{eqnarray}%
with $N=\sinh ^{2}r$, $M=e^{i\theta }\sinh r\cosh r$, where $r$ and $\theta$ are the squeezing parameter and phase of the two-mode squeezed vacuum field, which could be generated by a Josephson parametric amplifier (JPA)~\cite{sqzMW1}, by a Josephson mixer~\cite{sqzMW2}, or by the combination of a JPA and a microwave beamsplitter~\cite{sqzMW3,sqzMW4}. The magnon input noise operators $m_j^{in}$ are zero mean and correlated as follows
\begin{equation}
\left\langle  m_{j}^{in\dag}(t)m_{j}^{in}(t^{\prime }) ; m_{j}^{in}(t)m_{j}^{in\dag }(t^{\prime })  \right\rangle =(N_{m_{j}}; N_{m_{j}}+1)\delta(t-t^{\prime }), \,\,\,\, (j=1,2) 
\end{equation}
where $N_{m_{j}}=[\exp (\frac{\hbar \omega _{m_{j}}}{k_{B}T})-1]^{-1}$ is the equilibrium mean thermal magnon number of the $j$th mode, with $T$ the environmental temperature and $k_B$ the Boltzmann constant.

Since we are interested in the quantum correlation properties of the two magnon modes, we focus on the dynamic of the quantum fluctuations of the system. The fluctuations of the system are described by the following QLEs 
\begin{eqnarray}
\delta \overset{\cdot }{a}_{j} &=&-(\kappa _{a_{j}}+i\Delta _{a_{j}})\delta a_{j}-ig_{j}\delta m_{j}+\sqrt{2\kappa _{a_{j}}}S_{j}^{in},  \label{QLEs2} \\
\delta \overset{\cdot }{m}_{j} &=&-(\kappa _{m_{j}}+i\Delta _{m_{j}})\delta m_{j}-ig_{j}\delta a_{j}+\sqrt{2\kappa _{m_{j}}}m_{j}^{in}.  \notag
\end{eqnarray}
The above QLEs can be written in the quadrature form, with quadrature fluctuations defined as $\delta X_{j}=(\delta a_{j}+\delta a_{j}^{\dag })/\sqrt{2},\delta Y_{j}=i(\delta a_{j}^{\dag }-\delta a_{j})/\sqrt{2},\delta x_{j}=(\delta m_{j}+\delta m_{j}^{\dag })/\sqrt{2},\delta y_{j}=i(\delta m_{j}^{\dag }-\delta m_{j})/\sqrt{2}$ (similar definition for input noises $X_{j}^{in}, Y_{j}^{in}$ and $x_{j}^{in}, y_{j}^{in}$), which are
\begin{eqnarray}
\delta \overset{\cdot }{X}_{j} &=&-\kappa _{a_{j}}\delta X_{j}+\Delta
_{a_{j}}\delta Y_{j}+g_{j}\delta y_{j}+\sqrt{2\kappa _{a_{j}}}X_{j}^{in},
\label{QELs3} \\
\delta \overset{\cdot }{Y}_{j} &=&-\kappa _{a_{j}}\delta Y_{j}-\Delta
_{a_{j}}\delta X_{j}-g_{j}\delta x_{j}+\sqrt{2\kappa _{a_{j}}}Y_{j}^{in}, 
\notag \\
\delta \overset{\cdot }{x}_{j} &=&-\kappa _{m_{j}}\delta x_{j}+\Delta
_{m_{j}}\delta y_{j}+g_{j}\delta Y_{j}+\sqrt{2\kappa _{m_{j}}}x_{j}^{in}, 
\notag \\
\delta \overset{\cdot }{y}_{j} &=&-\kappa _{m_{j}}\delta y_{j}-\Delta
_{m_{j}}\delta x_{j}-g_{j}\delta X_{j}+\sqrt{2\kappa _{m_{j}}}y_{j}^{in}. 
\notag
\end{eqnarray}%
They can be cast in the matrix form 
\begin{equation}
\overset{\cdot }{u}(t)=Au(t)+n(t),  \label{MatrixForm}
\end{equation}
where $u(t)=[\delta X_{1},\delta Y_{1},\delta X_{2},\delta Y_{2},\delta x_{1,}\delta y_{1},\delta x_{2},\delta y_{2}]^{T}$, $A$\ is the drift matrix
\begin{equation}
A=\left( 
\begin{array}{cccccccc}
-\kappa _{a_{1}} & \Delta _{a_{1}} & 0 & 0 & 0 & g_{1} & 0 & 0 \\ 
-\Delta _{a_{1}} & -\kappa _{a_{1}} & 0 & 0 & -g_{1} & 0 & 0 & 0 \\ 
0 & 0 & -\kappa _{a_{2}} & \Delta _{a_{2}} & 0 & 0 & 0 & g_{2} \\ 
0 & 0 & -\Delta _{a_{2}} & -\kappa _{a_{2}} & 0 & 0 & -g_{2} & 0 \\ 
0 & g_{1} & 0 & 0 & -\kappa _{m_{1}} & \Delta _{m_{1}} & 0 & 0 \\ 
-g_{1} & 0 & 0 & 0 & -\Delta _{m_{1}} & -\kappa _{m_{1}} & 0 & 0 \\ 
0 & 0 & 0 & g_{2} & 0 & 0 & -\kappa _{m_{2}} & \Delta _{m_{2}} \\ 
0 & 0 & -g_{2} & 0 & 0 & 0 & -\Delta _{m_{2}} & -\kappa _{m_{2}} 
\end{array}
\right) ,  \label{driftmatrix}
\end{equation}
and $n(t)=[\sqrt{2\kappa _{a_{1}}}X_{1}^{in},\sqrt{2\kappa _{a_{1}}}Y_{1}^{in}, \sqrt{2\kappa _{a_{2}}}X_{2}^{in},\sqrt{2\kappa _{a_{2}}}Y_{2}^{in},\sqrt{2\kappa _{m_{1}}}x_{1}^{in},\sqrt{2\kappa _{m_{1}}}y_{1}^{in},\sqrt{2\kappa_{m_{2}}}x_{2}^{in},\sqrt{2\kappa _{m_{2}}}y_{2}^{in}]^{T}$. Since the dynamics of the system is linear and the input noises are Gaussian, the dynamical map of the system preserves the Gaussian nature of any input state. The steady state of quantum fluctuations of the system is therefore a continuous-variable four-mode Gaussian state, which is completely characterized by an $8\times 8$ covariance matrix (CM) $V$, defined as $V_{ij}(t)=\langle u_{i}(t)u_{j}(t^{\prime })+u_{j}(t^{\prime })u_{i}(t) \rangle/2$. When the system is stable, $t \to \infty $, the solution of $V$ can be obtained by directly solving the Lyapunov equation~\cite{DV07,Hahn}
\begin{equation}
AV+VA=-D,  \label{LyapEq}
\end{equation}
where $D$ is the diffuse matrix defined by $D_{ij}\delta (t-t^{\prime})=\langle n_{i}(t) n_{j}(t^{\prime })+n_{j}(t^{\prime })n_{i}(t) \rangle/2$. It can be written in the form of direct sum $D=D_{a}\oplus D_{m}$, where $D_{a}$ is related to the cavity modes, given by
\begin{equation}
D_{a}=\left( 
\begin{array}{cccc}
\kappa _{a_{1}}(2N+1) & 0 & \sqrt{\kappa _{a_{1}}\kappa _{a_{2}}}(M+M^{\ast
}) & i\sqrt{\kappa _{a_{1}}\kappa _{a_{2}}}(M^{\ast }-M) \\ 
0 & \kappa _{a_{1}}(2N+1) & i\sqrt{\kappa _{a_{1}}\kappa _{a_{2}}}(M^{\ast
}-M) & -\sqrt{\kappa _{a_{1}}\kappa _{a_{2}}}(M+M^{\ast }) \\ 
\sqrt{\kappa _{a_{1}}\kappa _{a_{2}}}(M+M^{\ast }) & i\sqrt{\kappa
_{a_{1}}\kappa _{a_{2}}}(M^{\ast }-M) & \kappa _{a_{2}}(2N+1) & 0 \\ 
i\sqrt{\kappa _{a_{1}}\kappa _{a_{2}}}(M^{\ast }-M) & -\sqrt{\kappa
_{a_{1}}\kappa _{a_{2}}}(M+M^{\ast }) & 0 & \kappa _{a_{2}}(2N+1)
\end{array}
\right) , 
\end{equation}
and $D_{m}$ is associated with the magnon modes, i.e., $D_{m}={\rm diag} \big[ \kappa _{m_{1}}(2N_{m_{1}}+1),\kappa_{m_{1}}(2N_{m_{1}}+1),\kappa_{m_{2}}(2N_{m_{2}}+1),\kappa_{m_{2}}(2N_{m_{2}}+1) \big]$.

Once the CM of the system is achieved, one can then calculate the degree of entanglement between the two magnon modes. We adopt the logarithmic negativity~\cite{LogNeg,GAJPA} to quantify the entanglement, which is defined as
\begin{equation}
E_N \equiv \max \big\{ 0, \, -\ln2\tilde\nu_- \big\},
\end{equation}
where $\tilde\nu_-\,\,{=}\,\min{\rm eig}|i\Omega_2\tilde{V}_{mm}|$ (with the symplectic matrix $\Omega_2=\oplus^2_{j=1} \! i\sigma_y$ and the $y$-Pauli matrix $\sigma_y$) is the minimum symplectic eigenvalue of the CM $\tilde{V}_{mm}={\cal P}_{1|2}{V_{mm}}{\cal P}_{1|2}$, where $V_{mm}$ is the $4\times 4$ CM of the two magnon modes, obtained by removing in $V$ the rows and columns of the two cavity modes, and ${\cal P}_{1|2}={\rm diag}(1,-1,1,1)$ is the matrix that implements partial transposition at the level of CMs~\cite{Simon}.

\begin{figure}[t]
\hskip-0.07cm\includegraphics[width=0.98\linewidth]{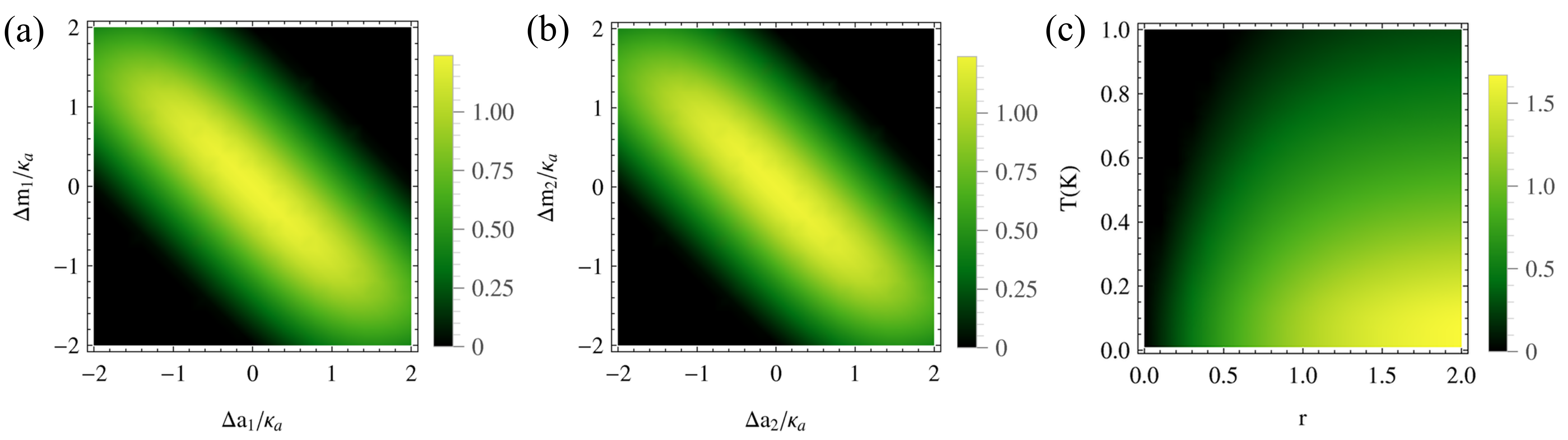} 
\caption{Density plot of the entanglement $E_{mm}$ between two magnon modes versus (a) $\Delta _{a_1}$ and $\Delta _{m_1}$,
with $\Delta _{a_2}=\Delta _{m_2}=0$, $r=1$, and $T=100$ mK; (b) $\Delta _{a_2}$ and $\Delta _{m_2}$, with $\Delta _{a_1}=\Delta _{m_1}=0$, $r=1$, and $T=100$ mK; (c) $r$ and $T$, with $\Delta _{a_{1,2}}=\Delta _{m_{1,2}}=0$. See text for the other parameters.}
\end{figure}

\section{Numerical results of magnon entanglement}
\label{Numeri}

In Fig. 2, we show the entanglement of two magnon modes versus various system parameters. Figure 2(a) and (b) show that the optimal situation for magnon entanglement is that in each cavity the cavity and magnon modes are resonant with the driving field, i.e., $\Delta _{a_j}=\Delta _{m_j}=0$ ($j=1,2$). This is consistent with the results in the study of transferring single-mode squeezing from microwave field to the magnon mode~\cite{Jie19PRA}. Physically this is easy to understand: owing to the linear cavity-magnon coupling, the resonant situation most efficiently transfers the quantum correlation from the input fields to the two magnon modes. Figure 2(c) shows that the entanglement increases with the squeezing of the input two-mode squeezed field and decreases with the temperature. Note that we have assumed the bandwidths of the input squeezed fields are larger than the cavity linewidths. In Ref.~\cite{sqzMW2}, a two-mode squeezed field with logarithmic negativity $E_N = 0.8$ (corresponding to squeezing $r=0.4$)~\cite{Note} and bandwidth of 12.5 MHz has been produced. With this we can achieve magnon entanglement $E_{mm}= 0.6$ at $T=100$ mK, and the entanglement survives up to $\sim$0.8 K. We have employed in Fig. 2 experimentally feasible parameters \cite{Strong2}: $\omega_{a_1}/2\pi =10$ GHz, $\kappa_a/2\pi =5$ MHz, $\kappa_{m}=\kappa _a/5$, $g_1=g_2=5\kappa _{a}$, and an optimal phase $\theta =0$. In Ref.~\cite{Strong2}, a YIG sphere with a diameter of 0.5 mm was used, which contains more than $10^{17}$ spins. Therefore, we consider that the magnon modes are at macroscopic scale and the entangled states of them can be referred to as macroscopic quantum states. Note that, for simplicity, we have taken equal cavity (magnon) decay rates, $\kappa_{a_{1}} = \kappa_{a_{2}} = \kappa_a$ ($\kappa_{m_{1}} = \kappa_{m_{2}} = \kappa_m$). However, the results obtained in this paper can be straightforwardly extended to the general case of unequal decay rates. The entanglement is in the steady state guaranteed by the negative eigenvalues (real parts) of the drift matrix $A$. Actually, the steady state is always guaranteed for realistic nonzero decay rates due to the specific form of the drift matrix.

\begin{figure}[b]
\hskip-0.8cm\includegraphics[width=0.7\linewidth]{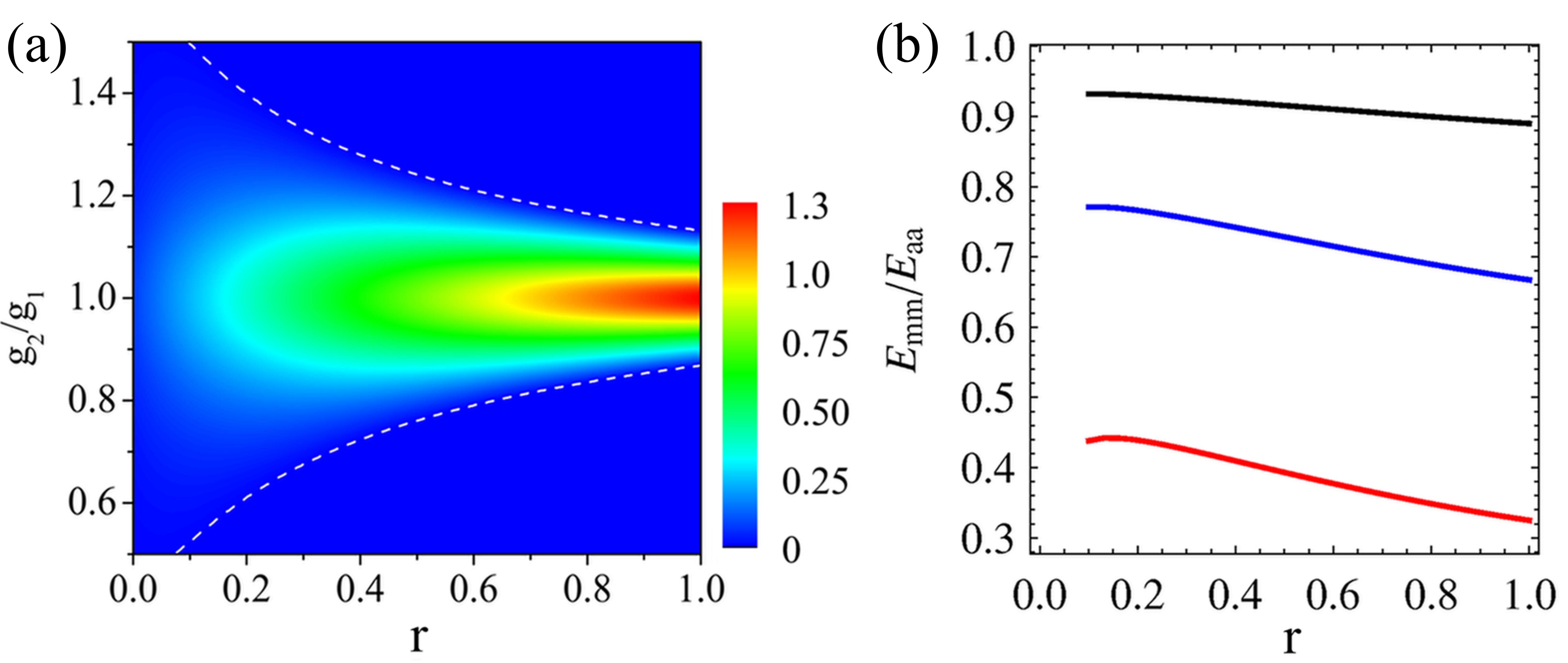} 
\caption{(a) Density plot of the entanglement $E_{mm}$ between two magnon modes versus $r$ and $g_2/g_1$, with $g_1=5\kappa _{a}$. The region between two dashed lines denotes $E_{mm}>0$. (b) $E_{mm}/E_{aa}$ versus squeezing $r$ for various couplings $g_1=g_2=0.5 \kappa_a, \kappa_a, 2\kappa_a$, corresponding to solid lines from bottom to top, respectively. The rest of the parameters are $\omega_{a_1}/2\pi =10$ GHz, $\Delta _{a_{1,2}}=\Delta _{m_{1,2}}=0$, $\kappa_a/2\pi =5$ MHz, $\kappa_{m}=\kappa _a/5$, $\theta =0$, and $T=100$ mK.}
\end{figure}

It would be interesting to investigate the effectiveness of the present scheme against the mismatch of the two couplings. In a similar proposal~\cite{Jaya}, a {\it single-mode} squeezed field is injected into one cavity to entangle two magnon modes. Specifically, the squeezed field is used to squeeze one collective quadrature of two magnon modes to violate specific inequalities thus demonstrating magnon entanglement. In there, identical coupling strengths are preferred to effectively implement the proposal. In contrast, our scheme uses a different mechanism: two magnon modes get entangled due to the quantum correlation transferred from a {\it two-mode} squeezed field, and this would overcome the limitation on the couplings. Indeed, as shown in Fig. 3(a), considerable entanglement is generated in a wide range of mismatch of the two couplings, and the situation of smaller squeezing $r$ is more tolerant to the mismatch. We also explore the entanglement transfer efficiency, reflected by the ratio $E_{mm}/E_{aa}$, from the two cavity modes to the two magnon modes. In Fig. 3(b), we plot $E_{mm}/E_{aa}$ as a function of squeezing $r$ for three cases of the couplings $g_1=g_2=0.5 \kappa_a$, $\kappa_a$, and 2$\kappa_a$. It is evident that in order to transfer the entanglement with high efficiency strong coupling $g_{1,2} > \kappa_a$ should be used. The fact that coupling strength as large as double cavity decay rate yields about 90\% transfer efficiency makes our scheme quite promising.

\section{Analytical solutions at optimal conditions}
\label{Analy}

In the preceding section, we have numerically shown optimal parameter regimes for the generation of sizable magnon entanglement. When the cavity and magnon modes are resonant with the input fields and the two couplings are strong and take close values, large magnon entanglement can be achieved which increases with the squeezing of the input fields. The entanglement is in the steady state and robust against environmental temperature. In this section, we explore more deeply the problem by providing analytical solutions under the above optimal conditions, where the steady-state CMs take relatively simple expressions.

The two cavity modes get entangled due to the injection of the two-mode squeezed vacuum field, which shapes the noise properties of quantum fluctuations of the cavity fields, i.e., the cavity input noise operators become quantum correlated. This can be clearly seen in the CM of the two cavity modes by setting the couplings $g_1=g_2=0$ (the magnons get decoupled) and $\Delta _{a_1}=\Delta _{a_2}=0$, which takes the following form
\begin{equation}\label{Vaa}
V_{aa} =\frac{1}{2}
\begin{pmatrix}
\cosh 2r  &  0  &  \sinh 2r  &  0   \\
  0  &  \cosh 2r  & 0  & - \sinh 2r   \\
\sinh 2r  &  0  &  \cosh 2r  &  0    \\
0    &  - \sinh 2r  &  0  & \cosh 2r   \\
\end{pmatrix} ,
\end{equation}
which is independent of cavity decay $\kappa_a$ and is exactly the CM of the two-mode squeezed vacuum state with squeezing $r$~\cite{GAJPA}. The logarithmic
negativity of such a state is 
\begin{equation}
E_{aa} = \max \Big\{ 0,  -{\rm ln} \big( \,{\rm min} \big\{  |\cosh r - \sinh r |^2 , \, |\cosh r + \sinh r |^2  \big\} \, \big)  \Big\} ,
\end{equation}
which increases with the squeezing $r$, as shown in Fig. 4.

\begin{figure}\label{fig4}
\hskip-1.0cm\includegraphics[width=0.4\linewidth]{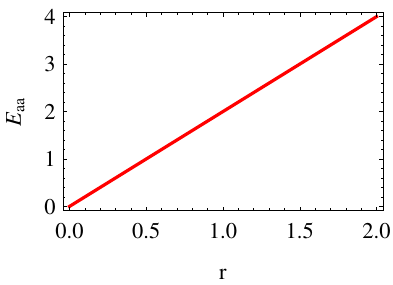}
\caption{Logarithmic negativity of the two cavity modes versus squeezing $r$ with $\Delta _{a_1}=\Delta _{a_2}=0$, $\theta=0$, and $g_1=g_2=0$.}
\end{figure}

The cavity-cavity entanglement is then partially transferred to the two magnon modes when the two couplings (beamsplitter type) are switched on, $g_1=g_2 = g >0$. The stationary CM of the two magnon modes can be achieved, which is
\begin{equation}\label{Vmm}
V_{mm} =\frac{1}{2(1{+}a)(a{+}b^2)}
\begin{pmatrix}
a (1{+}a{+}b^2) {+} b^2 \cosh 2r  &  0  &  - b^2 \sinh 2r  &  0   \\
  0  &  a (1{+}a{+}b^2) {+} b^2 \cosh 2r  & 0  &  b^2 \sinh 2r   \\
- b^2 \sinh 2r  &  0  & a (1{+}a{+}b^2) {+} b^2 \cosh 2r  &  0    \\
0    &  b^2 \sinh 2r  &  0  & a (1{+}a{+}b^2) {+} b^2 \cosh 2r   \\
\end{pmatrix} ,
\end{equation}
where $a=\kappa_m/\kappa_a$, $b=g/\kappa_a$, and we have assumed $N_m \simeq 0$, which is the case at low temperature $T <100$ mK for magnon frequencies $\omega_{m_j} \sim 10$ GHz. The logarithmic negativity of such a state is, however, too lengthy to be reported here. In Fig. 5, we show both the steady-state cavity entanglement and magnon entanglement as a function of $\kappa_m/\kappa_a$ and $g/\kappa_a$. It is clear that small magnon decay rates and large coupling rates, $\kappa_m \ll \kappa_a \ll g$, are preferred for obtaining large magnon entanglement, and as the couplings increase, the entanglement is gradually transferred from the two cavity modes to the two magnon modes. This is very much alike to the case of transferring single-mode squeezing from the cavity to the magnon mode~\cite{Jie19PRA}, where one quadrature of the magnon mode is optimally squeezed with large coupling and small magnon decay rate, $g \gg \kappa_a \gg \kappa_m$.

\begin{figure}[t]\label{fig5}
\includegraphics[width=0.8\linewidth]{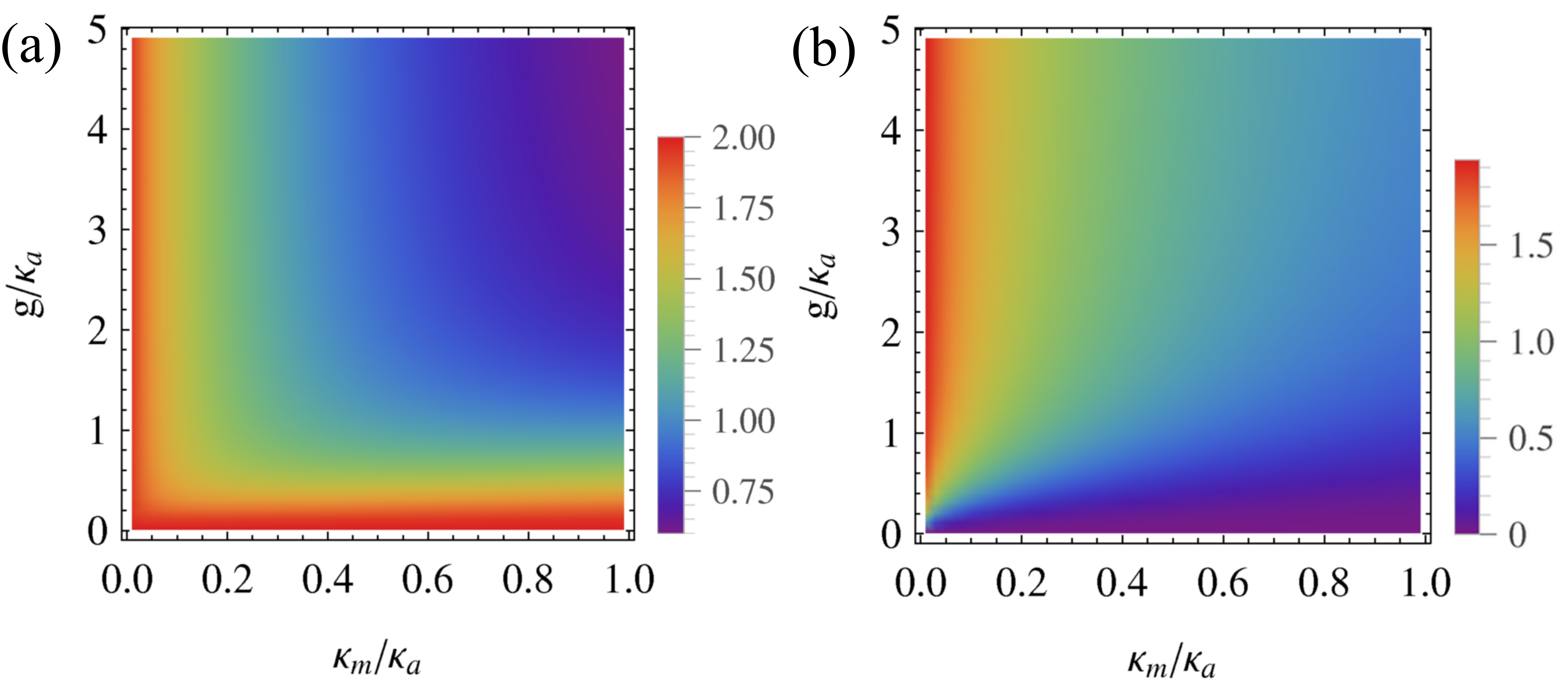}
\caption{Density plot of the entanglement (a) $E_{aa}$ of two cavity modes, (b) $E_{mm}$ of two magnon modes, versus $\kappa_m/\kappa_a$ and $g/\kappa_a$, with $\Delta _{a_{1,2}}=\Delta _{m_{1,2}}=0$, $\theta=0$, and $r=1$.}
\end{figure}

We note that in the steady state in each cavity the cavity and magnon modes never get entangled as their interaction is linear and is of beamsplitter type, $H_{int} = g \big( a m^{\dag }+a^{\dag }m \big)$~\cite{Jie18PRL,Jie19}. This is confirmed by the zero value logarithmic negativity. The CM of the cavity-magnon system in each cavity is given by
\begin{equation}\label{Vam}
V_{am} =\frac{1}{2(1{+}a)(a{+}b^2)}
\begin{pmatrix}
a b^2 {+} (a{+}a^2{+}b^2) \cosh 2r  &  0  &  0  &  - 2 a b \sinh^2 r   \\
  0  &  a b^2 {+} (a{+}a^2{+}b^2) \cosh 2r   &  2 a b \sinh^2 r  &  0   \\
0  &  2 a b \sinh^2 r   & a (1{+}a{+}b^2) {+} b^2 \cosh 2r  &  0    \\
 - 2 a b \sinh^2 r   & 0  &  0  & a (1{+}a{+}b^2) {+} b^2 \cosh 2r   \\
\end{pmatrix} ,
\end{equation}
and the logarithmic negativity can be written as $E_{am} \,{=} \,\max \big\{ 0, \cal{N} \big\} $, where $\cal{N}$ is a long expression and always nonpositive, as shown in Fig. 6, implying that the cavity and magnon modes are separable.

\begin{figure}
\includegraphics[width=0.4\linewidth]{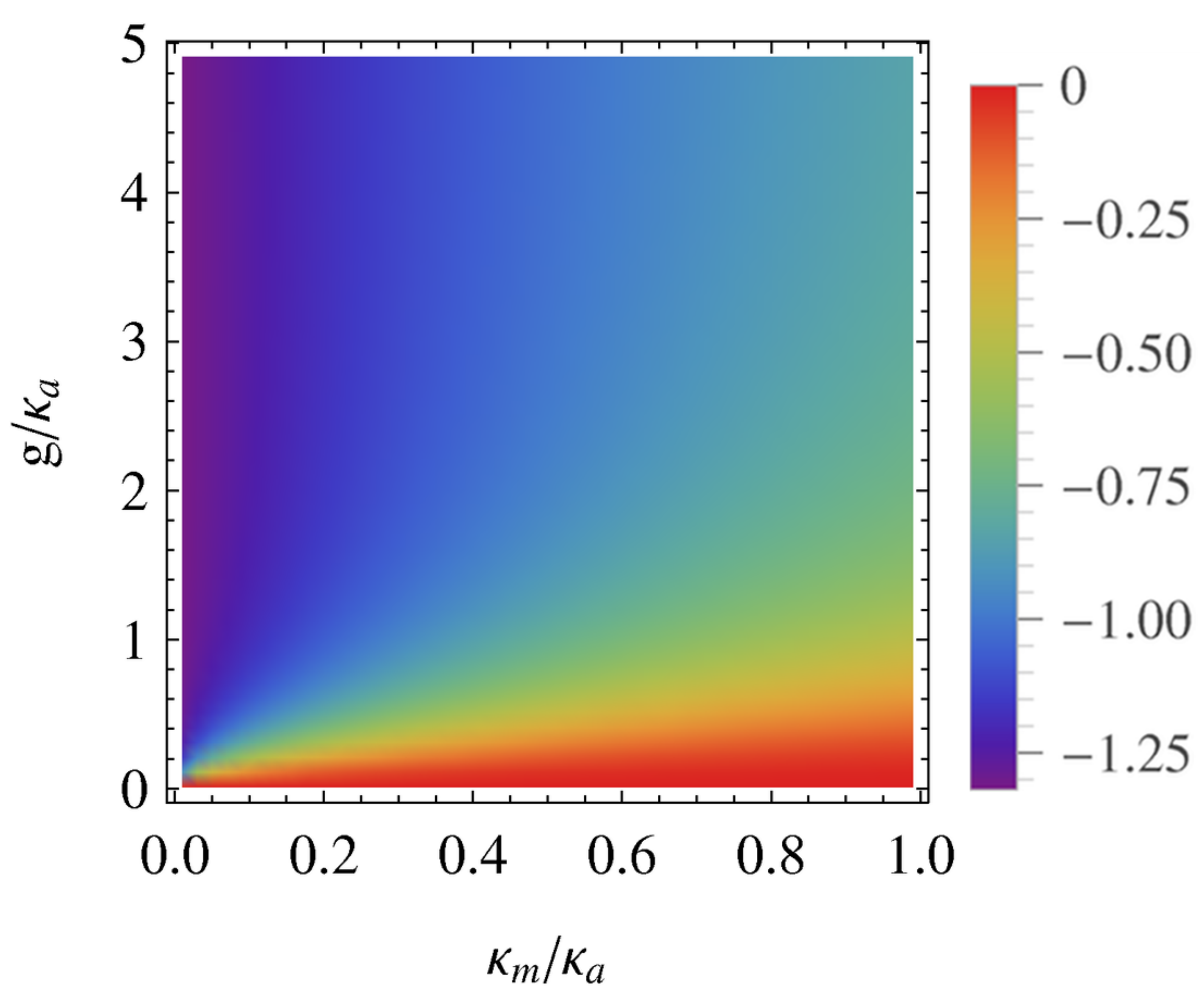}
\caption{Density plot of $\cal{N}$ as a function of $\kappa_m/\kappa_a$ and $g/\kappa_a$. The parameters are as in Fig. 5.}
\end{figure}

\section{Conclusions}
\label{Conc}

We have presented a scheme to prepare entangled states of two magnon modes in two massive YIG spheres via transferring quantum correlations from a two-mode squeezed microwave field. We have shown optimal parameter regimes for achieving strong magnon entanglement, and in particular, studied the effectiveness of the scheme towards the mismatch of two cavity-magnon couplings and analysed the entanglement transfer efficiency. Large coupling rates and small magnon decay rates with respect to cavity decay rates are preferred for the entanglement. We have shown that, with experimentally accessible two-mode squeezed source, strong magnon entanglement could be realized which survives up to hundreds of milliKelvin. Macroscopic entangled states of magnon modes are not only useful for fundamental studies of quantum-to-classical transition, decoherence theories at macroscopic scale~\cite{Bassi}, but can also be applied to quantum information processing based on magnonic systems~\cite{NakaRev} as valuable resources.

\begin{acknowledgments}

This research was supported by National Key Research and Development Program of China (Grants No. 2017YFA0304200 and No. 2017YFA0304202), National Natural Science Foundation of China (Grant No. 11674284), Zhejiang Provincial Natural Science Foundation of China (Grant No. LD18A040001), and the Fundamental Research Funds for the Center Universities (No. 2019FZA3005).

\end{acknowledgments}


\begin{thebibliography}{99}
\bibitem{Note}
Ref.~\cite{sqzMW2} reported entangled microwave fields of logarithmic negativity $E_N = 1.15$ with the base of 2. This corresponds to $E_N = 0.8$ in our definition with the base of natural constant $e$.
\end{thebibliography}
\end{document}